\documentclass[10pt,journal,compsoc]{IEEEtran}

\usepackage{comment}

\usepackage{amssymb}
\usepackage{pifont}
\usepackage[shortlabels]{enumitem}
\usepackage{multirow}
\usepackage{array}
\usepackage{makecell}
\usepackage{graphicx}
\usepackage{stackengine}
\newcommand\xrowht[2][0]{\addstackgap[.5\dimexpr#2\relax]{\vphantom{#1}}}
\usepackage[table, dvipsnames]{xcolor}

\ifCLASSOPTIONcompsoc

\usepackage[nocompress]{cite}
\else

\usepackage{cite}
\fi
\ifCLASSINFOpdf
\fi

\begin{document}

	\title {V2X in 3GPP Standardization:\\ NR Sidelink in Rel-16 and Beyond}

	\author{Mehdi Harounabadi,  Dariush Mohammad Soleymani, Shubhangi Bhadauria, Martin Leyh, Elke Roth-Mandutz
		\IEEEcompsocitemizethanks{\IEEEcompsocthanksitem M. Harounabadi is with the Department
			of Broadband and Networking, Institute for Integrated Circuits IIS , Erlangen,
			Germany.\protect\\
			E-mail: mehdi.harounabadi@iis.fraunhofer.de
		}
		\thanks{Copyright © 2021, IEEE, DOI: 10.1109/MCOMSTD.001.2000070 }}

	\markboth{This paper has been accepted to be published in "IEEE Communication Standards Magazine, March 2021"}%
	{Shell \MakeLowercase{\textit{et al.}}: Bare Advanced Demo of IEEEtran.cls for IEEE Computer Society Journals}

	\IEEEtitleabstractindextext{%
		\begin{abstract}
			
 			The 5G mobile network brings several new features that can be applied to existing and new applications. 
 			High reliability, low latency, and high data rate are some of the features which fulfill the requirements of vehicular networks. Vehicular networks aim to provide safety for road users and several additional advantages such as enhanced traffic efficiency and in-vehicle infotainment services. This paper summarizes the most important aspects of NR-V2X, which is standardized by 3GPP, focusing on sidelink communication. The main part of this work belongs to the 3GPP Rel-16, which is the first 3GPP release for NR-V2X, and the work/study items of the future Rel-17.
		\end{abstract}
		
		}

	\maketitle

	\IEEEdisplaynontitleabstractindextext

	\IEEEpeerreviewmaketitle

	\ifCLASSOPTIONcompsoc
	\IEEEraisesectionheading{\section{Introduction}\label{sec:introduction}}
	\else
	\section{Introduction}
	\label{sec:introduction}
	\fi

	\IEEEPARstart{V}{ehicular} networks have been studied for decades to provide communication among cars which are equipped with wireless interfaces. The main objective of the vehicular networks was limited to increase the safety of road users and improve the traffic efficiency of roads. 
	In recent years, more use cases have been planned for vehicular networks which are not limited only to road safety. The use cases mostly support the trending research and development in the automotive industry, i.e. autonomous driving. Furthermore, new technologies allow the users of vehicular networks to enjoy infotainment applications which require high throughput and low latency communication. 
	
	The 5th Generation (5G) mobile network, which has been released recently and is under development in many countries, is one of the candidate technologies to be employed in vehicular networks. There are some traditional studies and activities to use IEEE-based standards such as IEEE 802.11p \cite{jiang2008ieee}. However, the new features of 5G New Radio (NR) and its capability to support several different use cases make it a better choice for vehicular networks. 
	Sidelink (SL) communication, which was standardized in LTE by 3rd Generation Partnership Project (3GPP), is the key enabler of V2X communication. SL communication is direct communication between two User Equipments (UEs) without the participation of a base station in the transmission and reception of data traffic.

	This paper provides an insight into the general and specially Radio Access Network (RAN) aspects of SL communication in NR-V2X, which have been specified in 3GPP release 16 (Rel-16). Moreover, the paper reveals future aspects of NR-V2X which are currently under discussion for Rel-17 or planned for future studies beyond Rel-17. Similar work to this paper is \cite{5gv2x20} where the overall system architecture aspect of LTE/5G-V2X is reviewed. Other works such as \cite{ltev2x18} and \cite{ltev2x17} review and study LTE-V2X based on 3GPP Rel-14/15.  
	
	This paper is organized as follows. In Chapter 2, 3GPP standardization activities are explained. A short history of cellular-based V2X is described in Chapter 3 where different 3GPP releases for V2X are introduced. Chapter 4 explains the general aspects of the V2X application such as its use cases, coverage scenarios, and communication types. Chapter 5 describes the RAN aspects of NR SL communication with respect to the V2X application. It covers the overall, physical, and higher layer aspects of NR SL. Chapter 6 briefly compares LTE-V2X and NR-V2X and shows the enhancements in NR-V2X. An insight into the upcoming Rel-17 is provided by Chapter 7 which illustrates the trend in the standardization activities for V2X. Chapter 8 describes NR SL improvements beyond Rel-17. Finally, the paper is concluded in Chapter 9.

		\section{3GPP Standardization Activities}

	 3GPP standardization activities are organized into the following three Technical Specification Groups (TSG):
	\begin{itemize}
		\item \textbf{Radio Access Network (RAN)} TSGs consist RAN1, RAN2, RAN3, RAN4, RAN5, and RAN6. The responsibility of a RAN TSG is the definition of the functions, requirements and interfaces of the access network in the physical layer, layer 2, and layer 3 of the protocol stack.  Moreover, the conformance testing of UEs and base stations applying the defined solutions are the responsibilities of the RAN TSGs. 
		\item \textbf{Service and System Aspects (SA)} TSGs consist SA1, SA2, SA3, SA4, SA5, and SA6 that are responsible for the definition, development and maintenance of the overall system architecture, and service capabilities in the system. Moreover, they  support inter-TSGs coordination. 
		\item \textbf{Core Network and Terminals (CT)} TSGs consist CT1, CT3, CT4, and CT6 that are responsible for the definition of terminals, interfaces, and capabilities, the development of core network and its inter-connection with external networks for end-to-end networking.
	\end{itemize}
	Details of the tasks of TSGs are shown in Table \ref{table:3gpp}.
	The outcome of TSGs are documented as follows:
	\begin{itemize}
		\item Technical Report (TR): is the results of a Study Item (SI). An SI is the initial study (phase) of a topic to be considered for the specification procedure. 
		\item Technical Specification (TS): an SI becomes Work Item (WI) after an initial study and agreement. Then, the final agreements of a WI are covered in a TS.

	\end{itemize}
	There are different series of documents in 3GPP documentation relevant to the RAN aspects: 
	\begin{itemize}
		\item \textbf{36.xxx}: are documents with a focus on LTE (evolved UTRA), LTE-Advanced, LTE-Advanced Pro. radio technology
		
		\item \textbf{37.xxx}: are documents with a focus on multiple radio access technology aspects
		\item \textbf{38.xxx}: are documents with a focus on the radio technology beyond LTE (e.g. NR)
	\end{itemize}
	\begin{figure}[t]
		
		\centering
		\includegraphics[scale=0.4]{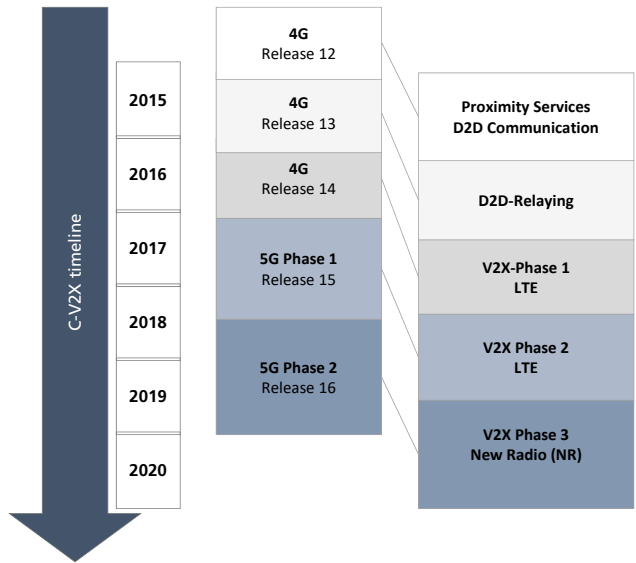}
		\caption{History of 3GPP releases. }
		\label{fig:flow}
	\end{figure}   
	\begin{table*}
	
		\begin{center}
			\caption{Main focus of different TSGs in 3GPP}	\label{table:3gpp}
			\begin{tabular}{|c | c | c|}
				\hline\rowcolor{cyan}
				\xrowht{25pt}
				
				\textbf{TSG RAN} &\textbf{ TSG SA} &\textbf{ TSG CT} \\
				\hline\xrowht{25pt}
				
				\shortstack{RAN WG 1 \\ Radio layer 1 specification} & \shortstack{SA WG 1 \\ Services} &\shortstack{CT WG 1 \\ MM/CC/SM (lu)} \\ 
				\hline\xrowht{25pt}
				\shortstack{RAN WG 2 \\ Radio layer 2, 3 specification} & \shortstack{SA WG 2 \\ Architecture}& \shortstack{CT WG 3 \\ Interworking with external networks}  \\
				\hline\xrowht{25pt}
				\shortstack{RAN WG 3 \\ Interface specification} & \shortstack{SA WG 3 \\ Security} &\shortstack{CT WG 4 \\ MAP/GTP/BCH/SS} \\
				\hline\xrowht{25pt}
				\shortstack{RAN WG 4 \\ Radio performance and protocol aspects} & \shortstack{SA WG 4 \\ Codec} & \shortstack{CT WG 6 \\ Smart card application aspects} \\
				\hline\xrowht{25pt}
				\shortstack{RAN WG 5 \\ Mobile terminal conformance testing} & \shortstack{SA WG 5 \\ Telecom management}&  \\
				\hline\xrowht{25pt}
				\shortstack{RAN WG 6 \\ Legacy RAN and radio protocol} & \shortstack{SA WG 6 \\ Mission critical applications} &  \\
				\hline
				
			\end{tabular}
		\end{center}
	\end{table*} 
	
	\section{History of Cellular V2X}
The 4th generation mobile network or Long Term Evolution (LTE) was defined from Rel-8 of 3GPP with an increased peak data rate, better support of Quality of Service (QoS), and  improved support of cell edge users starting in 2008.  Following is a short overview of 3GPP releases with respect to the V2X application: 
\begin{itemize}
	\item LTE D2D Rel-12: several new features were introduced in LTE-advanced (Rel-12) such as Device to Device (D2D) communication. D2D is the direct communication between two UEs which is an enabling feature for V2X communication. 
	\item LTE-V2X Rel-14 and 15:
	the first V2X communication in LTE was studied and standardized in Rel-14. Four modes of resource allocation were considered for D2D communication in LTE where 2 modes were specifically for V2X application (mode 3 and 4). Broadcast communication was the only  supported cast type in LTE-V2X. In LTE-V2X, all the coverage scenarios (in-coverage, out-of-coverage, and partial-coverage) were supported.
	The main intention in LTE-V2X was the support of use cases to provide road safety fulfilling the demanding reliability and latency requirements.
	
	\item NR-V2X Rel-16:
	it was the first specification of NR SL with a focus on V2X enhancing the reliability, latency, capacity, and flexibility. 
	The use cases for this release are not only limited to  road safety, but also targeted advanced use cases such as platooning, extended sensors, advanced driving, and remote driving. Different cast types such as unicast and multicast (groupcast) are also specified for V2X communication in addition to broadcast. Moreover, the NR-V2X in Rel-16 supports the Hybrid Automatic Receive reQuest (HARQ) to improve the reliability of the SL communication. 
	
\end{itemize}
The history of 3GPP releases from 12 to 16 is shown in Figure \ref{fig:flow}.

	\section{General Aspects of NR-V2X Application}
	As mentioned in the introduction, the new features of NR-V2X facilitate the realization of advanced use cases.   

	Several use cases are considered for NR-V2X in Rel-16 \cite{ts22.186}. The use cases have strict QoS requirements which should be provisioned by NR. In this section, the most important use cases, which are the baseline for the NR-V2X studies, are explained.   
	\begin{itemize}
		\item \textbf{Extended sensors}: 

		in this use case, the sensor data is exchanged among vehicles, pedestrians, and other road users through a vehicular network with  very low latency and high reliability to provide a holistic view of the environment in each vehicle. The holistic view can enhance any perception, planning, and control in road users.  				
		
		\item \textbf{Platooning}: 
 
		a platoon is like a chain of vehicles that follow each other with a reduced safety distance employing a low latency and highly reliable communication among them. The head of a platoon sends commands and receives feedback from the other members to control the platoon.

		\item \textbf{Remote driving}:
 
		this use case is mostly useful for driving in dangerous zones or areas with less variations of roads such as bus lines in public transportation. In the latter case, a bus may be driven through a cloud. 
		
		\item \textbf{Advanced driving}: it is the use case for fully or semi-automated driving where vehicles share their local sensing information with each other and a Road Side Unit (RSU), if present, to coordinate their trajectories and to make better maneuvers. Collision avoidance, safer driving, and improved traffic efficiency are some of the benefits of such a use case.
	
	\end{itemize}

	Different types of communication are considered in NR-V2X to fulfill the requirements of many use cases for such a network. A use case may apply one or a combination of the following communication types: 
	\begin{itemize}
		\item Vehicle-to-Vehicle (V2V)

		\item Vehicle-to-Infrastructure (V2I)

		\item Vehicle-to-Pedestrian (V2P)

		\item Vehicle-to-Network (V2N)
 
	\end{itemize}

	Also, the V2X communication for  improved road safety, increased traffic efficiency or even infotainment may take place in one of the following network coverage scenarios:  
	\begin{itemize}
		\item in-coverage: when the communicating UEs on SL (such as Pedestrian-UEs (P-UEs) or Vehicular-UEs (V-UEs)) are located within the coverage of a gNB.

		\item out-of-coverage: when all communicating UEs are out of the coverage of any gNB.

		\item Partial-coverage: when at least one of the communicating UEs is in-coverage and have a connection to a gNB.
 
	\end{itemize}	

	\section{NR-V2X Sidelink Communication Based on 3GPP Rel-16}
	\subsection{Overall RAN Aspects}
	\subsubsection{Protocol Stack}
\begin{figure*}[ht]
		
		\centering
		\includegraphics[scale=0.4]{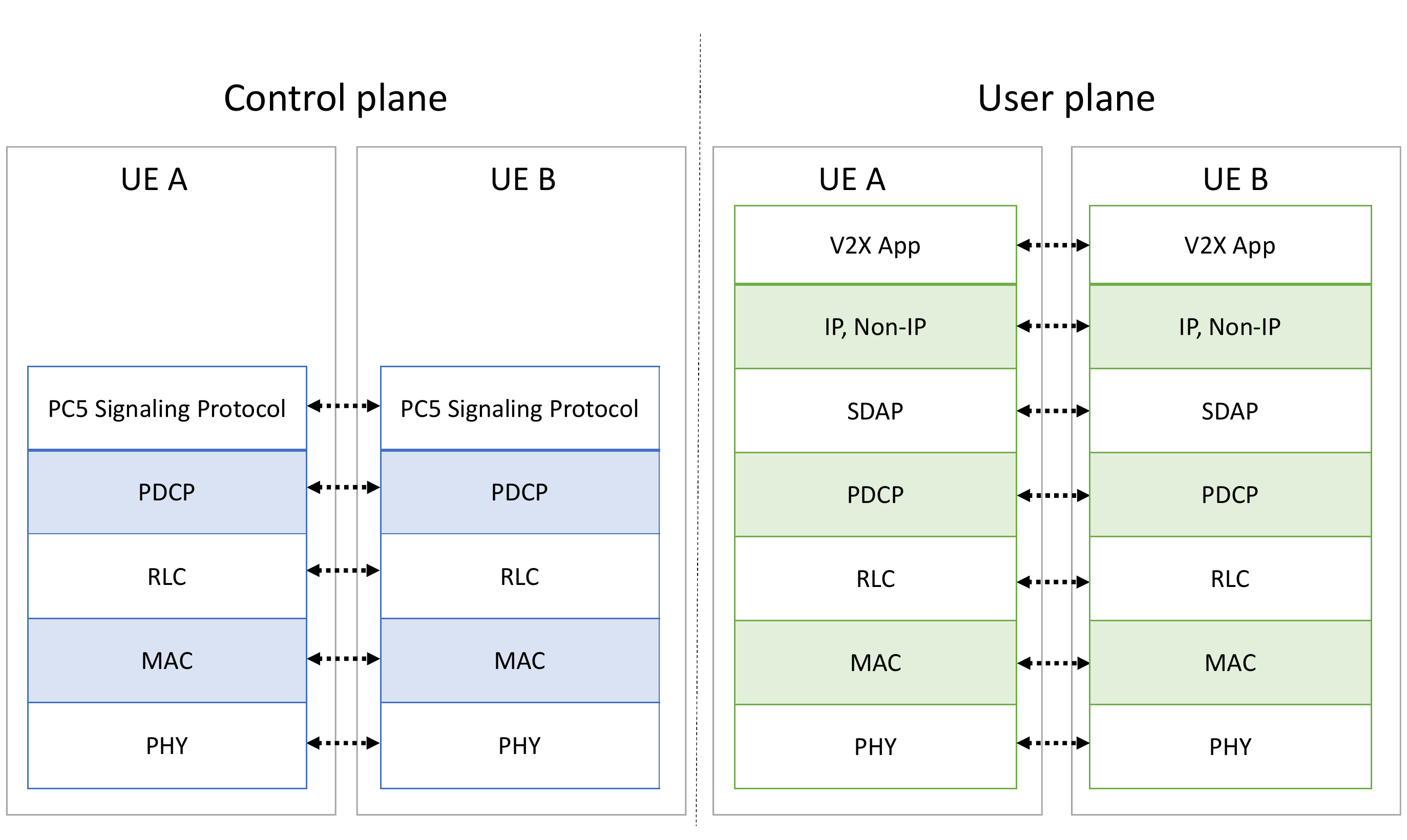}
		\caption{Control and user plane of SL communication in NR-V2X. }
		\label{fig:planes}
	\end{figure*}

 The protocol stack for NR SL in control and user planes is shown in Figure \ref{fig:planes}. The functionality of each layer is defined in \cite{ts38.300} and \cite{ts23287}. In this paper, the functionality and services provided by PHY, MAC, RLC, PDCP, SDAP, and RRC are explained briefly. The data and user planes are responsible for the transmission of user data and signaling, respectively.   
	\subsubsection{Cast Types}
	3GPP Rel-16 specifies three cast types for NR-V2X. The reason to support more cast types in comparison with LTE-V2X is to meet requirements of a wider range of use cases in vehicular networks. The supported cast types in NR-V2X are as follows:
	\begin{itemize}
		\item Unicast:  direct communication between a pair of UEs. 
		\item Broadcast: a single transmitter UE sends messages to be received by all UEs which may decode the message (within the radio transmission range of the transmitter UE). 
		\item Groupcast (multicast): a transmitter UE sends message(s) to a set of receivers which fulfill certain conditions (e.g being member of a group) 
	\end{itemize}
NR-V2X supports HARQ procedure for unicast and groupcast messages which can provide more reliability for theses traffic types. The SL HARQ is explained in Section \ref{sect:harq} 
	\subsubsection{Frequency ranges}
	For the NR SL in Rel-16, two frequency bands are defined as the operating bands of V2X on the PC5 interface \cite{ts38.1011}:
	\begin{itemize}
		\item 5.9 GHz band (n47)
		\item 2.5 GHz band (n38)
	\end{itemize}

	The supported channel bandwidth in both frequency bands are 10, 20, 30, and 40 MHz.
	
	The 5.9 GHz band is reserved for Intelligent Transport System (ITS) within Europe and worldwide for vehicular communication and is not limited to cellular-V2X (NR or LTE).  Basically, 30 kHz sub-carrier spacing with normal Cyclic Prefix (CP) is supported for NR SL.
	The 2.5 GHz band is used exclusively in particular regions for NR-V2X.
	Both frequency bands (5.9 and 2.5 GHz) are included in the Frequency Range 1 (FR1) ranging from 410 MHz to 7125 MHz. The initial design in Rel-16 is based on FR1.  
	The millimeter Wave (mmW) Frequency Range 2 (FR2) which ranges from 24.25 GHz to 52.6 GHz is not the primary focus of the studies in Rel-16.

\subsection{Physical Layer Aspects}
\subsubsection{Radio Resources}
The radio resources in NR are defined in time and frequency domains. SL has the same structure for radio frames, sub-frames, and slots as NR uplink/downlink \cite{ts38.211}. SL communication also supports different numerologies which result in shorter slot times. It is an enabling feature for use cases with a low latency requirement. 

\subsubsection{Physical Structure}

The NR SL physical layer is composed of several physical channels and signals \cite{ts38.211}.
The SL physical channels are a set of resource elements carrying information of higher layers of the protocol stack. The SL physical channels are  defined as follows:
\begin{itemize}
	
	\item Physical Sidelink Broadcast Channel (PSBCH): 
	it carries the SL-BCH transport channel where the Master Information Block (MIB) for SL is sent periodically (each 160 $ms$) and comprises system information for UE-to-UE communication (e.g. SL TDD configuration, in-coverage flag) \cite{ts38.331}. PSBCH is transmitted along with the Sidelink Primary Synchronization Signal/Sidelink Secondary Synchronization Signal (S-PSS/SSS) in the S-SSB (see synchronization signals).
	\item Physical Sidelink Feedback Channel (PSFCH): it is used to transmit the HARQ feedback from a receiver UE  to the transmitter UE on the SL for a unicast or groupcast communication.

	\item Physical Sidelink Shared Channel (PSSCH) and Physical Sidelink Control Channel (PSCCH):
	every PSSCH, which contains transport blocks i.e. user data traffic, is associated with a PSCCH. The PSCCH is transmitted on the same slot as PSSCH and contains control information about the shared channel. The Sidelink Control Information (SCI) is split into two stages. The 1st stage is sent on PSCCH, which is associated with a PSSCH, and the 2nd stage is sent over the corresponding PSSCH. 
	The content of SCI in the 1st and 2nd stages are as follows: 
		\begin{itemize}
		\item 1st stage  SCI
			\begin{itemize}
				\item Priority
				\item Frequency resource assignment
				\item Time resource assignment
				\item Resource reservation period
				\item DMRS pattern
				\item 2nd-stage SCI format
				\item Modulation and coding scheme
				\item Reserved
				\item Beta\_offset indicator
				\item Number of DMRS port

			\end{itemize}
		\item 2nd stage SCI
		\begin{itemize}
			\item HARQ process ID
			\item New data indicator
			\item Redundancy version
			\item Source ID
			\item Destination ID
			\item CSI request
			
		\end{itemize}
	\end{itemize}  
	 PSSCH is transmitted in consecutive symbols of a slot. The start symbol and the number of symbols to transmit the PSSCH are configured by the higher layer, i.e. MAC \cite{ts38.214}. A PSSCH cannot be transmitted in the same symbols which are configured for the transmission of PSFCH and also the last symbol of the slot, which is configured as a place holder for a guard symbol. 
\end{itemize}
SL physical signals are classified into reference signals and synchronization signals. The following SL  physical signals are specified in Rel-16:
\begin{itemize}
	
	\item Demodulation Reference Signals (DM-RS): they are used for PSCCH, PSSCH, and PSBCH as reference signals for  demodulation of messages in a receiver.
	\item Channel State Information Reference Signal (CSI-RS): it is a reference signal used for channel state estimation/sounding and reporting between a transmitter and a receiver UE.
	\item Phase Tracking Reference Signal (PT-RS): it is used as a reference signal for phase noise compensation.
	\item Sidelink Primary/Secondary Synchronization Signal (S-PSS/S-SSS): together with PSBCH are parts of the Sidelink Synchronization Signal Block (S-SSB) and used for the SL synchronization.

\end{itemize}

\subsubsection{Modulation Scheme and Numerology in NR Sidelink}
NR SL supports different OFDM numerologies i.e. sub-carrier spacing. The supported sub-carrier spacings for NR SL are 15, 30, 60 kHz for FR1 and 60, 120 kHz for FR2 with the same mapping to the frequency ranges and cyclic prefixes (such as normal and extended) as in NR downlink/uplink. 
NR SL is based on Cyclic Prefix OFDM (CP-OFDM) which supports QPSK, 16/64/256-QAM modulation schemes to provide high reliability or high throughput based on the requirements of the application and the channel state.

Table \ref{table:numerology} shows the transmission numerology in NR and the type of cyclic prefix for each numerology. 
\begin{table}

	\begin{center}
		\caption{Transmission numerology in NR}	\label{table:numerology}
		\begin{tabular}{|c | c | c|}
			\hline
			$\mu$ & Sub-carrier spacing (kHz)  & Cyclic prefix\\
			\hline \hline
			0 & 15 & Normal \\
			\hline
			1 & 30 & Normal \\
			\hline
			2 & 60 & Normal, Extended \\
			\hline
			3 & 120 & Normal \\
			\hline
		\end{tabular}
	\end{center}
\end{table}
	\subsubsection{Slot Format}
\begin{figure}[t]
	
	\centering
	\includegraphics[scale=0.3]{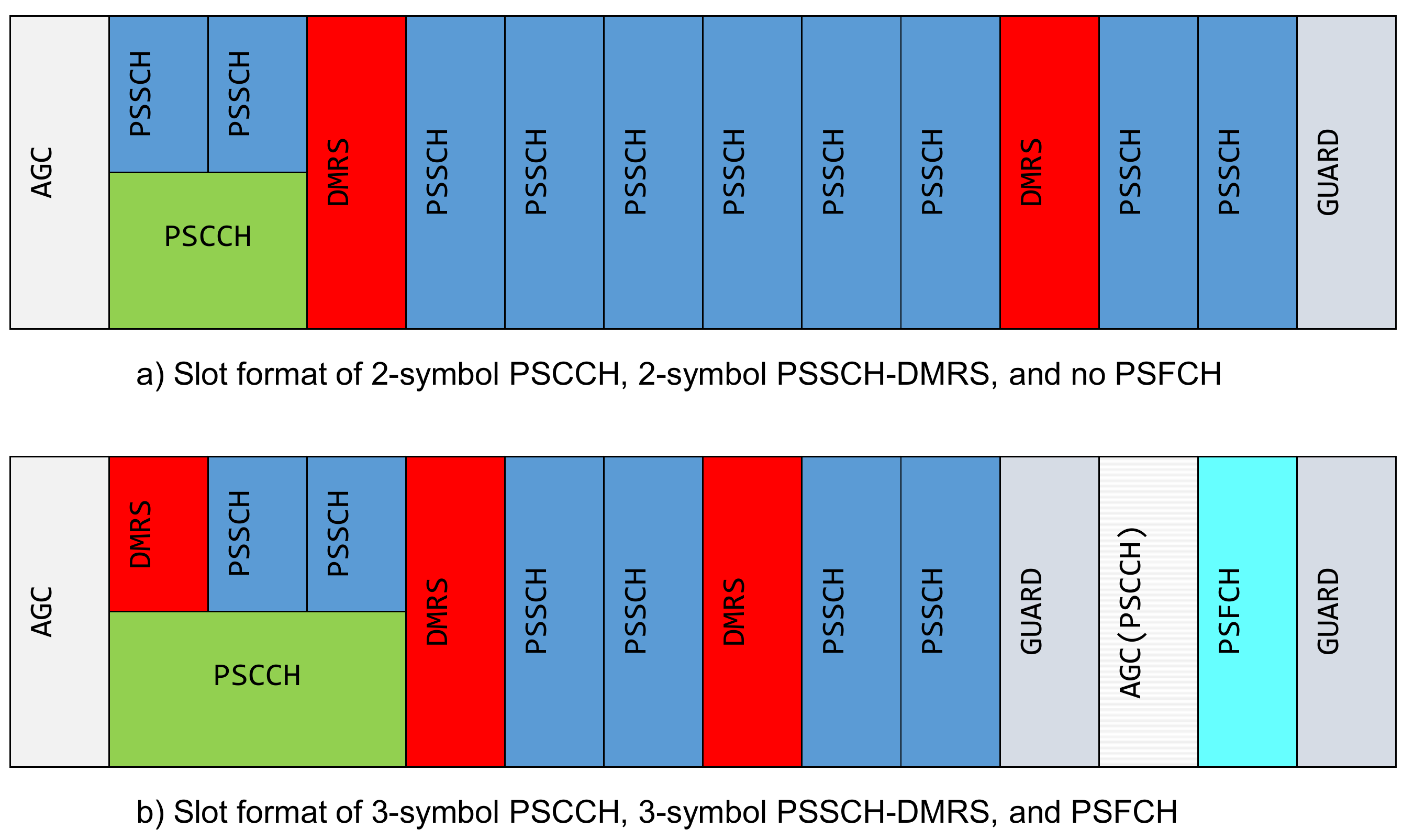}
	\caption{Slot format. }
	\label{fig:slot}
\end{figure}

The slot format of NR SL is the same as the NR slot format in downlink/uplink. Figure \ref{fig:slot} illustrates two types of slot formats and the order of symbols in a slot.  Each slot can include PSSCH, PSCCH, PSFCH, AGC and guard symbols. Automatic Gain control (AGC) and guard symbols are sent as specific symbols. AGC symbols are used for level control in a SL receiver whereas guard symbols are used as guard periods for switching between SL reception and transmission.
Guard symbols are placed as immediate symbols after PSSCH, PSFCH, or S-SSB.

	\subsubsection{Radio Resource Allocation}
	Three important concepts in the NR-V2X  radio resource allocation are described in this section. The concepts are bandwidth part, resource pool, and the modes of resource allocation. \\

	\textbf{Bandwidth part:} a BandWidth Part (BWP) is defined in \cite{ts38.211} for NR as "a contiguous set of physical resource blocks, selected from a contiguous subset of the common resource blocks for a given numerology on a given carrier". This definition is also valid for the SL BWP to specify the implementation of Radio Frequency (RF) hardware chain  in a UE. 	
	A UE is configured with only one active SL BWP when it  is connected to a gNB and  while it is in idle mode or out-of-coverage.  
    The BWP configuration contains the sub-carrier spacing (numerology) in the SL and it is associated with frequency ranges for SL communication. 
    The transmission and reception of a UE in SL should take place within a single SL BWP.\\
	
	\textbf{Radio resource pool:} a resource pool limits the radio resources for PSCCH and PSSCH since they cannot be transmitted in all Resource Blocks (RBs) and slots of NR or even the frequency span of the NR SL. 
	
	The concept of resource pool is also applied in autonomous resource allocation of UEs (see mode 2 resource allocation) where resources are selected based on a sensing procedure on a specific resource pool.
	
	In the frequency domain, a resource pool is divided into sub-channels which are consecutive and non-overlapping Physical Resource Blocks (PRB) where the number of PRBs is $\geq$ 10.
	The size of a resource pool is configured by higher layers or through signaling by a gNB.
	 	
	Transmission and reception resource pools are configured in a UE separately. 
	
	Additionally, an exceptional resource pool is configured for each UE for some exceptional cases such as Radio Link Failure (RLF) or handover where the UE has not any stable configuration of the transmission resource pool, but V2X communication is required.
	The exceptional resource pool is only for temporary usage. Resource allocation in an exceptional resource pool is based on a random selection of resources. 
	UEs receive the configuration of exceptional resource pools through the broadcasting of the serving cell or some dedicated signaling. 
	All UEs are mandated to monitor the exceptional resource pool in addition to the reception resource pool to enable the communication  for UEs which use these resources in exceptional cases.

	\textbf{Modes of resource allocation:} two modes of resource allocation are specified for NR SL communication which are applied in the V2X application. The modes are different mainly in terms of the entity which makes the decision on resources to be used for SL communication. 
	\begin{enumerate}
		\item Mode 1: in this mode, the resources are allocated by a gNB for in-coverage UEs. 
		There are two types of configured grants for mode 1:
	 \begin{itemize}
	 	\item Type 1: a sidelink configured grant is configured/released for the UEs via RRC signaling and can be used immediately. This type of resource can be used when a UE detects a channel impairment, e.g. beam failure, before an exceptional resource pool to be used.
	 	\item Type 2: a gNB grants the users a permission to activate or deactivate the configured resources through a Downlink Control Information (DCI) signaling.	
	 \end{itemize}
		
		\item Mode 2: is an autonomous resource selection by a UE based on a sensing procedure. 
		The sensing takes place in a pre-configured resource pool. UEs can select resources for transmission and re-transmission if the resources are not in use by other UEs with higher priority traffic. A UE may occupy resources for an appropriate amount of time until a re-selection event is triggered.
		The resources can be reserved by a UE for different purposes such as blind re-transmissions and HARQ feedback-based re-transmissions where the information about the reservation of resources is indicated in SCI. 
		
		In this mode of resource allocation, the UE performs  continuous sensing and when a resource selection event is triggered (e.g. arrival of a transport block) in the UE, it considers its recent sensing results between 1100 $ms$ and 100 $ms$ prior to the trigger time for resource selection. The 1100 $ms$ sensing window is beneficial to identify the reserved resources by other UEs for periodic traffic, which are sent in the 1st stage SCI on PSCCH, and the 100 ms sensing results are particularly useful for non-periodic traffic.   
		A UE, which performs sensing, measures the SL-RSRP of either PSCCH or PSSCH. This measurement is beneficial for a UE to select appropriate resources and avoid interference to any existing communication.
		 
		The selection of resources by a UE for its (re-) transmission is based on the sensing results and on a resource selection window. The selection window starts shortly after the trigger of resource selection in a UE (i.e. the arrival of a transport block) and the maximum window size is bounded to the remaining delay budget of the transport block.  
		The UE excludes the resources where the measured RSRP is higher than a threshold and considers them as occupied if the traffic priority in the measured resources is higher than its traffic priority. Otherwise, the UE may select the occupied resources if it has higher priority traffic. In this way, higher priority traffic can occupy resources even if they have been already reserved by other UEs. 
	
		After the exclusion of reserved resources from the selection window, a UE selects resources randomly from the best set of the remaining resources, i.e. 20\% of the best resources with a measured RSRP less than the (pre-) configured threshold and based on the traffic priority. If the remaining resources after the exclusion procedure are less than 20\% of the all resources in the selection window, then the UE relaxes the RSRP threshold (by 3dB) until it has at least 20\% (or 35, 50\%  based on traffic priority) of all resources in the selection window for resource allocation.   
		The final selection of resources is done in the MAC layer of a UE receiving a set of unoccupied resources from the physical layer. The selection of resources in the MAC is based on a random procedure. 
		
		The selected resources are not periodic and a UE may reserve up to three resources adding an indication in SCI. The reserved resources can be independent in time and frequency. 
		Figure \ref{fig:mode2} shows the sensing window, selection window, and the trigger time $n$ for resource allocation. There are two processing times ($T_{proc0}$ and $T_{proc1}$) before and after the trigger time $n$ which refer to the required time in the physical and MAC layers for processing and inter-layer information exchange.

		Shortly before the start of a transmission on a selected resource, a UE may re-evaluate its selection for late SCI receptions, after the sensing window, to be sure that the selected resource is still suitable for its transmission.

		\begin{figure*}[ht]
			
			\centering
			\includegraphics[scale=0.5]{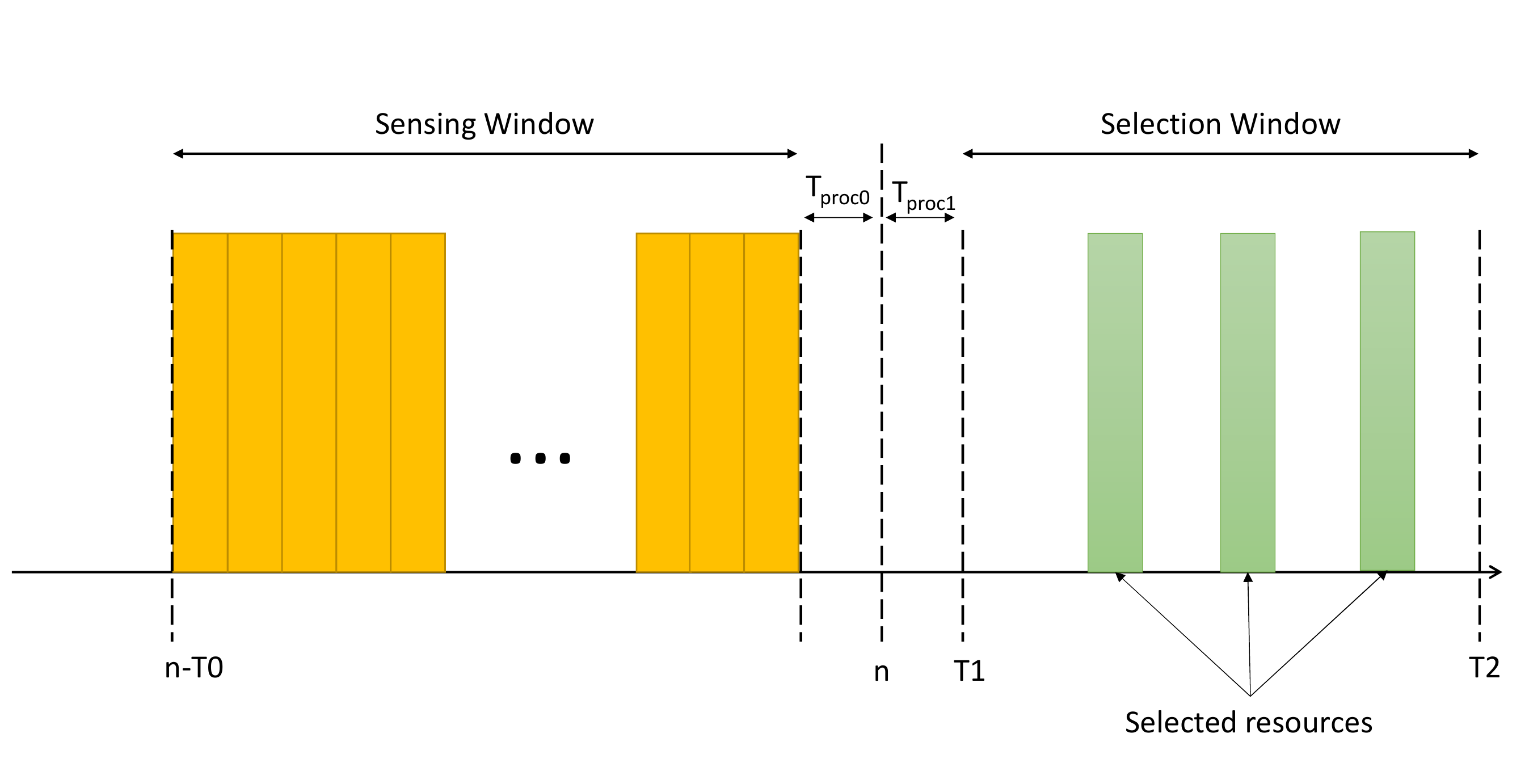}
			\caption{Mode 2 resource allocation in NR. }
			\label{fig:mode2}
		\end{figure*}

		After the selection of resources by a UE in mode 2, there could be several reasons for resource re-selection. One of the reasons can be the reception of a reservation from another UE in the vicinity with a higher priority traffic than the traffic which the UE is intending to transmit. 

	\end{enumerate}

	\subsubsection{Congestion control}
	The congestion control mechanism is applied in NR-V2X for mode 2 resource allocation and is similar to LTE-V2X. Two types of measurements are used in a UE to detect and mitigate the effect of congestion. Channel Busy Ratio (CBR) and Channel occupancy Ratio (CR) are the measurements in a UE to control the congestion when mode 2 resource allocation is applied. CBR  is calculated in each subframe as the portion of busy resources in a resource pool in the recent 100 ms. A resource is considered as busy if the received energy level in the resource is higher than a given threshold. CBR measurement results reflect the congestion in the medium. Another measurement is CR which reflects the occupancy of resources by a UE. CR is the number of sub-channels which a UE occupies and will occupy in a time window of 1000 ms. 
	
	In LTE, along with CBR and CR measurements, two parameters are applied in the congestion control mechanism. The parameters are based on the priority and reliability requirement of a packet and passed to the physical layer of a UE by higher layers. ProSe Per-Packet Priority (PPPP) reflects the priority of a packet and ProSe Per-Packet Reliability (PPPR) advocates the reliability requirement of a packet. PPPP and CBR are considered in a UE to limit the CR of the UE based on the priority of a packet. Moreover, transmission parameters such as the Modulation Coding Scheme (MCS) and maximum transmission power of the UE might be adapted based on the PPPP and CBR. 
	
	The congestion control in NR-V2X has few differences in comparison with LTE-V2X. Instead of PPPP and PPPR of a packet, only the priority of a packet is defined by higher layers and passed to the physical layer which is indicated in the 1st stage SCI. CR and transmission parameters are adapted based on the relative priority of the traffic in UEs and CBR. To react to the fast variations in the congestion of resources, NR-V2X defines shorter intervals for the calculation of CBR and CR. The intervals are 1 or 2 $ms$ in NR-V2X compared to 4 $ms$ in LTE-V2X.

	\subsubsection{QoS in Physical Layer}
	QoS control is not provided only in the physical layer. However, some mechanisms are applied, i.e. preemption, to support QoS in the physical layer. In NR-V2X, QoS requirements in the physical layer are indicated by a priority field in the 1st stage SCI. The priority field is 3 bits, which points out 8 different QoS classes (levels) mapping to corresponding QoS classes in the higher layers. 
	A UE prioritizes its transmission corresponding to the priority  of the higher layer QoS control mechanism. 
	Then, it reserves a set of resources for its transmission according to the resource selection procedure \cite{ts38.214}. 
	
	The UE may request a preemption of already reserved resources by other UEs from the higher layer when the resources indicated in the received 1st stage SCI (from other UEs) have been excluded from the identified resources during the resource selection procedure. To make such a request, the following conditions should be met:
	\begin{itemize}
		\item The preemption feature is enabled and the priority of  transmission in a  UE is higher than the priority indicated in the received 1st stage SCI (transmission of other UEs).
		\item The preemption feature is not enabled, but the priority indicated in the received 1st stage SCI (by other UEs) is lower than a specific threshold and is lower than the priority of the UE transmission.
	\end{itemize}

	It is also possible to configure a resource pool with a preemption function. In this case, a UE re-selects all the resources which it has already reserved in a particular slot if another UE in the proximity indicates the transmission of higher priority traffic in any of the reserved resources.

	\subsubsection{Synchronization}
	UEs need to be synchronized to have a common view from the time domain. This is an essential issue for UEs to access the medium based on a synchronized clock. There are different sources (references) of synchronization for UEs as follows:
	\begin{itemize}
		\item GNSS
		\item gNB or eNB
		\item Another UE transmitting SideLink Synchronization Signal (SLSS) which is called SynchRef UE
		\item The internal clock of a UE
	\end{itemize} 
	The first two synchronization sources (i.e. GNSS and gNB/eNB) are considered as the highest-quality sources. 
	
	The SyncRef UEs are synchronized either directly or indirectly to GNSS or an eNB/gNB. The indirect synchronization means that the GNSS or eNB/gNB are 1 or 2 hops away from the UE. 
	As a final option, when a UE is unable to be synchronized with GNSS or eNB/gNB directly or through a SynchRef UE, it uses its internal clock to transmit S-SSB.

	\subsection{Higher Layer Aspects}
 
This section explains briefly the services and functionalities of higher layers in the NR-V2X SL protocol stack. 
\subsubsection{Sidelink MAC}\label{sect:harq}
 
Functionalities provided by the SL Medium Access Control (MAC) are as follows: 

\begin{itemize}
	\item Radio resource selection
	\item Packet filtering
	\item Priority handling between uplink and SL in a UE
	\item SL channel state information reporting
\end{itemize}
It should be noted that the above-mentioned functionalities for SL are additional to the functionalities which are provided by a MAC entity in a UE such as (de-) multiplexing, HARQ procedure and logical channel prioritization. 

Hybrid Automatic Repeat reQuest (HARQ) is an extended feature for SL communication in NR which is provided by the MAC layer. The HARQ procedure is supported for unicast and groupcast communication in NR-V2X to provide reliability. 
There are two types of HARQ procedures in NR SL as follows:
\begin{enumerate}
	\item ACK/NACK based feedback: it is similar to Uu (downlink/uplink) HARQ scheme where the feedback is sent based on the success or failure in the reception of the whole transport block.
	\item NACK-only feedback: it is particularly for groupcast services. It reduces the number of resources needed when a high number of receiver UEs exist and need to send feedback at the same time. An example of this type of HARQ feedback is an extended sensors scenario where several UEs will receive information from a transmitter UE and the re-transmission occurs only if any of the receiving UEs fails to receive the sensory information correctly. A communication range, which is a radius of reception for the message, is defined by higher layers for the message. If any UE within the communication range of a message fails to decode the message, the receiver UE sends a NACK feedback.  
\end{enumerate}
PSFCH carries one bit of HARQ feedback which is sent by a receiver to a transmitter UE. 
In mode 1 of resource allocation  where resources are allocated by the gNB, UEs inform the gNB about a HARQ feedback using Physical Uplink Shared CHannel (PUSCH) or Physical Uplink Control CHannel (PUCCH) to assist it in the allocation of resources for re-transmissions.

SL transport channels are Service Access Points (SAPs) between SL physical and MAC layers. SL MAC can access  the services provided by the physical layer through the following transport channels:
\begin{itemize}
	\item Sidelink Shared Channel (SL-SCH)
	\item Sidelink Broadcast Channel (SL-BCH)
\end{itemize}

SL MAC provides data transfer services on logical channels. To accommodate different types of data transfer services, multiple types of logical channels are defined i.e. each supporting transfer of a particular type of information. Each logical channel type is defined by the type of information that is transferred in it. The SL logical channels which are SAPs between SL MAC and RLC are as follows:
\begin{itemize}
	\item Sidelink Broadcast Control Channel (SBCCH): a channel for broadcasting SL system information from a UE to other UE(s).
	\item Sidelink Control Channel (SCCH): a channel for transmission of control information (i.e. PC5-RRC and
	PC5-S messages) from a UE to other UE(s).
	\item Sidelink Traffic Channel (STCH): a channel for transmission of  user information from a UE to other UE(s).
\end{itemize}

\subsubsection{Sidelink RLC}
SL Radio Link Control (RLC) supports three transmission modes:
\begin{itemize}
	\item Transparent Mode (TM): is used for SBCCH. 
	\item Unacknowledged Mode (UM): is the only transmission mode for groupcast and broadcast. However, this mode can also be used for unicast.  
	\item Acknowledged Mode (AM): is used only for unicast. 

\end{itemize}
Services provided by SL RLC are sequence numbering, error correction through ARQ, segmentation, and reassembly. Some of the services are only provided for a specific transfer mode.    
\subsubsection{Sidelink PDCP}
Some of the services and functions of the Packet Data Convergence Protocol (PDCP) are as follows: 
\begin{itemize}
	\item Header (de)compression
	\item (De)ciphering
	\item Duplication/discarding duplicates
	\item In/out of order delivery
	\item Maintenance of a PDCP sequence number
	\item Integrity protection and verification
\end{itemize}

The services are supported for SL with some restrictions as follows:
\begin{itemize}
	\item Out of order delivery is supported only for unicast.
	\item Duplication is not supported in SL.
\end{itemize}

\subsubsection{Sidelink SDAP}
The Service Data Adaptation Protocol (SDAP) provides the mapping service in the user plane between a QoS flow and a SL Data Radio Bearer (SL-DRB). There is only one SDAP entity per destination for a unicast, groupcast, and broadcast communication on the SL.
\subsubsection{Sidelink RRC}
The Radio Resource Control (RRC) provides the following services in SL:
\begin{itemize}
	\item Transfer of a PC5-RRC message between two UEs: the messages could be configuration (such as SL-DRB configuration) or UE capability information. A UE capability is a piece of information that indicates different parameters that can be supported w.r.t. different layers. RRC provides Information Elements (IEs) for transferring such information between UEs.
	\item Maintenance and release of a PC5-RRC connection between two UEs: a PC5 connection is a logical connection between two UEs after the establishment of a PC5 unicast link. A UE may have one or multiple connections to another UE. 
	\item Detection of SL RLF for a PC5-RRC connection based on an indication from MAC or RLC: in the case of SL RLF, the UE releases the resources of the PC5 RRC connection immediately and sends an indication to the upper layer. For RRC\_CONNECTED UEs, the UE also informs the network upon RLF detection.
\end{itemize}

	\section{Comparison of LTE-V2X and NR-V2X}
	In this chapter, the main differences between LTE-V2X and NR-V2X are explained. As more use cases are considered for NR-V2X, new features are to support the requirements of the use cases. Some of the differences between the two generations of V2X are as follows \cite{ts37.985}: 
	\begin{itemize}
		
		\item Radio resource allocation:  in LTE mode 1 and 2 were defined for the D2D communication which are not related to V2X in Rel-12/13. Mode 3 and 4 were added in Rel-14 for V2X resource allocation.
		In NR-V2X, mode 1 and 2 are specified for resource allocation. 
		\item Cast types: broadcast communication was the only supported cast type in LTE-V2X which was limiting its use cases to broadcast safety messages such as Cooperative Awareness Message (CAM) messages. However, NR-V2X supports unicast and groupcast which makes it suitable for a wide variety of applications such as platooning and extended sensors.
		\item SL HARQ procedure: is an additional feature in Rel-16 for NR-V2X to enhance the reliability of unicast and groupcast communication using re-transmissions based on a receiver feedback. LTE-V2X does not provide such a feature in SL since it supports only broadcast communication in SL. 
		\item Flexible slot structure: supporting different sub-carrier spacing i.e. 15, 30, 60, 120 kHz for SL, NR-V2X can deliver shorter slots and lower transmission latency in comparison with LTE-V2X, which only supports 15 kHz sub-carrier spacing in SL.
		\item Physical channels: PSFCH in NR SL carries HARQ feedback from a receiver to a transmitter UE. The HARQ mechanism is not supported in LTE-V2X and therefore there is no feedback channel in LTE SL. 
		\item Protocol stack: the SDAP sub layer is added to the user plane in the SL V2X protocol stack to provision flow-based QoS.

	\end{itemize}

	\section{NR-V2X Sidelink Communication in 3GPP Release 17}
	In Rel-16, SL was specified for the first time based on NR, with a focus on V2X application. In addition, other services, i.e. public safety are not precluded, if their service requirements can be met. The need for further enhancements was identified as Rel-16 SL cannot support all requirements and operational scenarios of V2X, public safety, and other commercial use cases. 
	In 3GPP RAN plenary meeting (RAN\#86) for Rel-17 in December 2019, a set of SIs and WIs were agreed.  
	The SIs and WIs aim at several further enhancements which focus on wider coverage, increased reliability, reduced latency, and power saving for battery-based UEs. 
	
	Based on the RAN plenary agreement, a WI on SL enhancements to be studied in RAN1, a SI on SL relaying to be studied in RAN2, and finally a quite limited, high level SI on SL positioning to be studied in RAN plenary were defined. 

In the following, the WIs and SIs in Rel-17 are explained shortly.	
 
	\subsection{Sidelink Enhancements}
Rel-17 SL enhancements \cite{RP-201385} includes the major steps forward for the NR SL communication. This WI started in RAN1  focusing on the following two major enhancements:
\begin{enumerate}
	\item Power saving for battery-powered UEs: typically UEs carried by pedestrians and cyclists  using V2X application are battery-powered. For UEs operating on Rel-16, NR SL is assumed to be “always-on”, which is applicable for vehicular-based UEs with quasi unlimited batteries. However, for all types of battery-powered UEs, e.g. Vulnerable Road Users (VRUs) using V2X or UEs using public safety services, power saving is essential.
	\item Reliability and latency improvement:  to meet the Ultra Reliability Low Latency Communication (URLLC) requirements of SL services even under critical network conditions, e.g. high load or poor channel conditions which cannot be met by the Rel-16 NR SL. 
	
\end{enumerate}

The following enhancements can be assigned to one of the above-mentioned directions of this WI:
\begin{itemize}
 \item Enhancement on autonomous resource allocation: to reduce UEs power consumption. The baselines are the solutions which have been implemented in LTE, i.e. random resource selection and partial sensing.
 \item Inter-UEs coordination: to improve reliability and latency in autonomous resource allocation. The idea is that a UE can offer some of its (reserved or allocated) resources to another UE for resource selection and transmission.
 
 \item Sidelink discontinuous reception (DRX):  it is especially important for power saving UEs, to allow them to go to possibly extended sleep mode on the SL. Besides the definitions of ON/OFF duration on the SL, the alignment of the SL DRX wake-up time among UEs communicating with each other and  between the SL and the Uu DRX are essential.  
\end{itemize}

	\subsection{Sidelink Relaying}
	The main objective in the SL relaying SI is to extend the coverage of the SL communication and the network (cell). Moreover, the power efficiency and enhanced QoS support are also additional goals for non-V2X applications and services.
	The following two types of relaying are investigated in this SI\cite{RP-193253}:
	\begin{itemize}
		\item UE-to-Network: it is required to extend the coverage of a cell and provide reachability for cell-edge users (or out-of-coverage users) which need to reach the Packet Data Network (PDN).
		\item UE-to-UE: in Rel-16, only single-hop SL communication is supported.  For out-of-coverage scenarios, the single-hop Sl communication may not be sufficient to ensure SL coverage. Therefore, a UE-to-UE relay can extend SL coverage.  
	\end{itemize}

	The aspects to be studied in this SI are relay (re-) selection, relay/remote UE authorization, QoS provisioning, service continuity, and security mechanisms with respect to the relay node architecture (L2 or L3). In addition, a discovery procedure to discover a relay node will be studied.

	\subsection{Sidelink Positioning}

	This SI was defined for multiple V2X and public safety use cases with accurate positioning requirements. For example,  relative longitudinal position accuracy of less than 0.5 $m$ for UEs is required in a platooning use case \cite{ts22.186}. 
	The SI investigates use cases, requirements, and scenarios for UEs in-coverage, partial-coverage, and out-of-coverage. In detail,  the SI has two main objectives. First, to identify positioning use cases and requirements for V2X and public safety based on the existing 3GPP work and the input from the industry. Second, to identify potential deployment and operation scenarios. It is expected that the results will be used as input for new RAN SI and WI in Rel-18.

	\section {NR Sidelink improvements beyond Release 17}
	The definition of NR SL was started in Rel-16 and is under study to be enhanced in Rel-17. However, several new use cases are still expected. The use cases are not only related to V2X and public safety but also related to industrial communication, e.g. SL between machines and robots. 
	
	SL positioning is the missing functionality in Rel-17 to fulfill the strict requirements on precise positioning. The high level study in Rel-17  can only be considered as a starting point. For example, relative positioning using the SL, which is retrieving the distance between vehicles, is seen as a major important functionality by many companies since several releases. Therefore, SL positioning is expected as one of the major WIs in Rel-18.
	
	Another topic in 3GPP SA1, is a study on vehicle-mounted base station relays that provide service to UEs inside the vehicle or in the vicinity of the vehicle. 
	
	In addition, QoS awareness with respect to resource allocation and power efficiency may need to be considered as a highly relevant aspect, especially in highly loaded networks to ensure successful communication with high QoS requirements. A further topic under discussion in SA2 \cite{tr23776} is QoS-aware power-efficient SL communication for pedestrian UEs.
	
	Finally, SL communication may be further improved using Artificial Intelligence (AI)  or Machine Learning (ML). For example, resource allocation considering environmental conditions (e.g. road and weather conditions), road-specific conditions (e.g. current and maximum allowed speed) as well as radio conditions as input to intelligent algorithms, is expected to improve the reliability, latency, and overall throughput of future communication on  SL.

	\section{Conclusion}
	This paper summarized the most important aspect of NR-V2X in 3GPP release 16. NR-V2X introduces additional features to LTE-V2X such as different sub-carrier spacing for lower latency communication, SL HARQ mechanism to increase reliability, support for unicast and groupcast, faster reaction to congestion in SL for V2X applications. In addition, it gives an insight into the general, physical layer and higher layer aspects of NR Rel-16 RAN for SL communication. Moreover, the new topics in 3GPP Rel-17, which are under study and discussion for V2X applications, were described. Rel-17 of NR SL aims to provide lower latency, higher reliability, extended coverage, and reduced power consumption for battery-based UEs for the V2X application.

	\ifCLASSOPTIONcompsoc

	\ifCLASSOPTIONcaptionsoff
	\newpage
	\fi

	\bibliographystyle{IEEEtran}

	\begin{IEEEbiographynophoto}{Mehdi Harounabadi}
		Mehdi Harounabadi finished his bachelor's and master's in computer
		engineering in 2006 and 2009, respectively. He received his Ph.D. in
		computer science with great praise at the Technical University of
		Ilmenau. Since 2020, he has been working as a research scientist at
		Fraunhofer Institute in Erlangen and a lecturer at the University of Applied
		Sciences in Ingolstadt. Besides, he is a delegate in 3GPP
		standardization and 5G Automotive Association (5GAA) meetings. His
		research interests are 5G-V2X, autonomous driving, and AI-enabled
		mobile networks. 

	\end{IEEEbiographynophoto}

	\begin{IEEEbiographynophoto}{Dariush Mohammad Soleymani}
    Dariush Mohammad Soleymani received his bachelor’s and master’s in Telecommunication in 1999 and 2011. He is a Ph.D. student at the electrical engineering faculty of the Technical University of Ilmenau.  Since 2015, he has been working on device-to-device and vehicular communications in 4G and 5G networks. Also, he has been an RF radio planner and optimization engineer for 15 years at the leading operators and Nokia corporation in different countries. Since 2019, he is a research associate and 3GPP RAN WG1 delegate at the Fraunhofer Institute in Erlangen. His research focuses on resource allocation, artificial intelligence, and energy-efficiency. 
	\end{IEEEbiographynophoto}

	\begin{IEEEbiographynophoto}{Shubhangi Bhadauria}
	Shubhangi Bhadauria received her B.E. in electronics and telecommunication from Bhilai Institute of Technology (BIT), Bhilai, India (2013), and M.Sc. in communication and multimedia engineering from Friedrich Alexander University (FAU), Erlangen, Germany (2016). She is currently, working towards her Ph.D. with FAU and Fraunhofer IIS, Erlangen, Germany. She is also working on 3GPP standardization on 5G-V2X. Her research interests include vehicular communications, radio resource management, and machine learning. 
	\end{IEEEbiographynophoto}


	\begin{IEEEbiographynophoto}{Martin Leyh}
		Martin Leyh graduated from Friedrich-Alexander University Erlangen-Nürnberg in 1997 with a diploma degree in electrical and electronics engineering. He joined the Fraunhofer Institute for Integrated Circuits in 1997 and is currently chief engineer of the broadband and broadcast department. His research interests are in the area of mobile communication systems and standards, artificial intelligence and machine learning, and rapid system design and prototyping.
	
	\end{IEEEbiographynophoto}

	\begin{IEEEbiographynophoto}{Elke Roth-Mandutz}
	Elke Roth-Mandutz joined Fraunhofer IIS in 2017 and is currently the program manager for 5G standardization with a focus on the topics NR sidelink, and V2X. In addition, she is involved in the evolution of 5G technologies. She has also been involved in the development of wireless access technologies for 2G (GSM) and 3G (UMTS) at Alcatel-Lucent, followed by a scientific career at the TU Ilmenau. She has a Ph.D. on the topic of self-organizing networks for energy-saving in LTE and holds numerous patents and publications in the field of wireless communications.
	
\end{IEEEbiographynophoto}

\end{document}